# Dielectric Fluid in Inhomogeneous Pulsed Electric Field

**M.N. Shneider[1*] and M. Pekker[2]**

[1]*Department of Mechanical and Aerospace Engineering, Princeton University, Princeton, NJ 08544, USA*
[2]*Drexel Plasma Institute, Drexel University, 200 Federal Street, Camden, NJ 08103, USA*

## Abstract
We consider the dynamics of a compressible fluid under the influence of electrostrictive ponderomotive forces in strong inhomogeneous nonstationary electric fields. It is shown that if the fronts of the voltage rise at a sharp, needle-like electrode are rather steep (less than or about nanoseconds), and the region of negative pressure arises, which can reach values at which the fluid loses its continuity with the formation of cavitation ruptures. If the voltage on the electrode is not large enough or the front is flatter, the cavitation in the liquid does not occur. However, a sudden shutdown of the field results in a reverse flow of liquid from the electrode, which leads to appearance of negative pressure, and, possibly, cavitation.

## Introduction
A study of the behavior of liquid dielectrics in electric fields has a long history, which was started by Faraday [1]. It is known that dielectric fluids in strong non-uniform electric field are influenced by electrostrictive ponderomotive force [2-4]. As a result, fluid tends to be set in motion and moves into the regions with the strongest field. However, if the voltage rise time on the sharp electrode is very steep, the fluid does not have enough time to come into motion due to inertia. Consequently, the ponderomotive forces cause significant electrostrictive tensile stress. In other words, a region of so-called negative pressure arises in the fluid. It is known (see, eg [5-7]) that at a certain threshold of negative pressure the fluid loses its continuity, resulting in developing of the cavitation ruptures. If the rise time of the electric field is long enough and the liquid has time to be set in motion, the flow arising by the action of electrostriction forces reduces the value of negative pressure down to below the cavitation threshold and the discontinuities do not occur.

In the recent years, the development of breakdown in liquid dielectrics in the sub-nanosecond and nanosecond pulsed nonuniform electric fields have been extensively studied experimentally (see, for example, recent works [8,9] and references therein). As was shown in the paper [10], the appearance of cavitation raptures (pores) in the fluid due to the electrostrictive negative pressure formation, in particular, may promote the development of electrical breakdown at these conditions, which we will not consider in this paper. Note, that ponderomotive electrostrictive effects in the dielectric fluid also are possible in the non-uniform mean square field of the laser radiation. Thus, in [11] example was shown that shape of liquid droplets can be modified by the volumetric electrostrictive forces arising in the vicinity of inhomogeneous laser beam.

*shneyder@princeton.edu

## Physical Model and Equations

In this paper, we consider the dynamics of a compressible fluid under the influence of electrostrictive ponderomotive forces in strong inhomogeneous nonstationary electric fields, which are sub-critical to the breakdown development.

In general, the volumetric force acting on the dielectric fluid in nonuniform electric field are determined by the Helmholtz equation [2-4]:

$$\vec{F} = e\delta n\vec{E} - \frac{\varepsilon_0}{2}E^2\nabla\varepsilon + \frac{\varepsilon_0}{2}\nabla\left(E^2\frac{\partial\varepsilon}{\partial\rho}\rho\right), \qquad (1)$$

where the first term is the force acting on non-neutral fluid with the density of free charges $e\delta n$, the second and third terms are volumetric densities of ponderomotive forces, $\varepsilon_0$ is the vacuum dielectric permittivity, $\rho$ is the liquid density, $\vec{E}$ is the electric field. The second term in (1) is associated with the force acting on an inhomogeneous dielectric, and the third term corresponds to electrostrictive forces in a non-uniform electric field associated with the tensions within the dielectric.

In the absence of breakdown, we can disregard the forces acting on the free charges and with the inhomogeneity of the liquid. In this case, the body force acting on the liquid dielectric is reduced to

$$\vec{F} \approx \frac{\varepsilon_0}{2}\nabla\left(E^2\frac{\partial\varepsilon}{\partial\rho}\rho\right) \approx \frac{\varepsilon_0}{2}\left(\frac{\partial\varepsilon}{\partial\rho}\rho\right)\nabla E^2 \qquad (2)$$

where for nonpolar dielectrics, as follows from the Clausius-Mosotti formula, [12]:

$$\frac{\partial\varepsilon}{\partial\rho}\rho = \frac{(\varepsilon-1)\cdot(\varepsilon+1)}{3} \qquad (3)$$

and for polar dielectrics (water)

$$\frac{\partial\varepsilon}{\partial\rho}\rho = \alpha\varepsilon \qquad (3')$$

where $\alpha \le 1.5$ is the empirical factor for most of the studied polar dielectric liquids, including water, [13,14].

The stretching internal stresses, which are associated with the action of the volumetric forces (1) or (2), can lead to formation of micro ruptures (cavitation) in the fluid. The possibility of rupture of fluid under the influence of electrostriction is mentioned in [10].

In accordance with the nucleation theory [5], the critical tension for the fluid rupture is given by:

$$p_c = p_{sat} - \left( \frac{16\pi\sigma^3}{3kT \ln\left(NB / J\right)} \right), \tag{4}$$

Here $p_{sat}$ is the vapor pressure of liquid at a given temperature $T$, $k$ is the Boltzmann constant, $\sigma$ is the surface tension coefficient, $J$ is the nucleation rate equal to the density of vapor bubbles of a critical size appearing per 1 second, $B = 10^{11}$ s is the kinetic coefficient which weakly depends on the temperature, and $N$ is the density of molecules of the fluid. In practice, the experimental limit stretching tension is much smaller than that predicted by the theory of homogeneous nucleation (equation (4)) [5,6]. Experiments [7] show that at initially normal conditions (room temperature and atmospheric pressure) water ruptures at a negative pressure of about 0.15 MPa while slowly stretching. However, at the rapid stretching water preserves continuity for higher negative pressures [7]. Measurements show that $p_c$ depends on many parameters, including the degree of purity of the fluid and the presence of dissolved gases and dust particles. Experimental data for water are in the range between ~6 and 50 Mpa [6].

We are studying the dynamics of dielectric liquid (water) in a pulsed inhomogeneous electric field in the approximation of compressible fluid dynamics within the standard system of equations of continuity of mass and momentum [15]:

$$\frac{\partial \rho}{\partial t} + \nabla\left(\rho \vec{u}\right) = 0 \tag{5}$$

$$\rho\left( \frac{\partial \vec{u}}{\partial t} + \left(\vec{u} \cdot \nabla\right)\vec{u} \right) = -\nabla p + \vec{F} + \eta_d\left(\Delta \vec{u} + \tfrac{1}{3}\nabla(\nabla \cdot \vec{u})\right) \tag{6}$$

and the Tait equation of state for "compressible" water [16,17]:

$$p = \left(p_0 + B\right)\left( \frac{\rho}{\rho_0} \right)^{\gamma} - B \tag{7}$$

$$\rho_0 = 1000 \text{ kg/m}^3, \quad p_0 = 10^5 \text{ Pa}, \quad B = 3.07 \cdot 10^8 \text{ Pa}, \quad \gamma = 7.5$$

Here $\rho$ is the fluid density, $p$ is the pressure, $\vec{u}$ is the velocity, $\eta_d$ is the dynamic viscosity. Body force in (6) acting on the polar fluid is given by (2), (3’)

$$\vec{F} = \frac{\varepsilon_0}{2}\nabla\left(E^2 \frac{\partial \varepsilon}{\partial \rho}\rho\right) \approx \frac{\alpha\varepsilon\varepsilon_0}{2}\nabla E^2 \tag{8}$$

Due to the gradient form of the ponderomotive force (8), it is convenient to write in (6):

$$-\nabla p + \vec{F} = -\nabla\left(p - \frac{1}{2}\left(\frac{\partial \varepsilon}{\partial \rho}\rho\right)\varepsilon_0 E^2\right). \tag{9}$$

That is, as noted above, the ponderomotive electrostriction force is reduced to an additional negative pressure stretching the liquid. Heating of fluid is neglected, since the change in fluid kinetic energy during the pulse is considerably smaller than its internal energy.

A standard set of boundary conditions for equations (5,6) in a cylindrical coordinate system with the axis along the electrode axis was applied:

$$\begin{array}{lll}
u_r\big|_{\Gamma} = 0 & u_z\big|_{\Gamma} = 0 & \left.\dfrac{\partial \rho}{\partial r}\right|_{\Gamma} = 0 \\
u_r\big|_{r=0} & \left.\dfrac{\partial u_z}{\partial r}\right|_{r=0} = 0 & \left.\dfrac{\partial \rho}{\partial r}\right|_{r=0} = 0 \\
\left.\dfrac{\partial u_r}{\partial r}\right|_{r=R} = 0 & u_z\big|_{r=R} = 0 & \left.\dfrac{\partial \rho}{\partial r}\right|_{r=R} = 0 \\
u_r\big|_{z=L} = 0 & \left.\dfrac{\partial u_z}{\partial z}\right|_{z=L} = 0 & \left.\dfrac{\partial \rho}{\partial z}\right|_{z=L} = 0
\end{array} \tag{10}$$

Here, R, L are the boundaries of the computational domain; $\Gamma$ is the electrode surface (Fig. 1). We assumed the no-slip condition (the fluid velocity at the electrode goes to zero) and the continuity of the fluxes of the density and momentum on the boundaries of the computational domain.

For the nanosecond time scales, the boundary layer $d$ is much smaller than the radius of the electrode's tip, $r_{el} \sim 1-10$ μm, (the characteristic size of the strong electric field in the fluid), i.e. $d \approx \sqrt{\nu t} \approx 10^{-7}$ m << $r_{el}$, where $\nu = \eta_d / \rho$ is the kinematic viscosity. For water at $T$ = 293 K: $\eta_d \approx 10^{-3}$ Pa·s; $\nu \approx 10^{-6}$ m$^2$/s. The characteristic time to establish the boundary layer, whose size is comparable to the radius of curvature of the electrode, $r_{el}$: $\tau \approx r_{el}^2 / \nu = 10^{-6} - 10^{-4}$ sec, is by 2-4 orders longer than the front rise time for the high-voltage pulse at the electrode. Therefore, the terms related to the viscosity can be neglected in equation (6).

Since we consider the processes in highly inhomogeneous field in the vicinity of a sharp needle-like electrode, which can be represented as a prolate ellipsoid, it is convenient to solve the equations for compressible fluid in prolate spheroidal coordinates, $\eta, \mu$ [18]. In this case, the equipotential surfaces $\Phi$ coincide with surfaces $\eta = const$.

$$\Phi(\eta) = \Phi_0 \frac{\ln(\coth(0.5\eta))}{\ln(\coth(0.5\eta_0))} \qquad (11)$$

Here, $\Phi_0 = U(t)$ is the potential on the electrode, the value $\eta_0 = a \cdot \coth(\xi)$ corresponds to an equipotential surface that coincides with the electrode; $\xi$ is the ratio of the semiaxes of the prolate ellipsoidal electrode;

$$a = r_{el} \cosh(\eta_0)/\sinh^2(\eta_0) \qquad (12)$$

is the focal distance, $r_{el}$ is the radius of curvature of the electrode tip

The relations linking the cylindrical coordinates with the prolate spheroidal coordinates in the axisymmetric case are as follows

$$\begin{aligned} r &= a \cdot \mathrm{sh}(\eta) \cdot \sin(\mu) \\ z &= a \cdot \mathrm{ch}(\eta) \cdot \cos(\mu) \\ \eta_0 &\le \eta < \infty \\ 0 &\le \mu \le \pi/2 \end{aligned} \qquad (13)$$

The corresponding Jacobian is:

$$H = \left| \frac{\partial(r,z)}{\partial(\eta,\mu)} \right| = a^2 \left( \cosh(\eta) - \cos^2(\mu) \right) \qquad (14)$$

The electric field in the variables $\eta, \mu$ is directed only along the $\eta$ axis and is equal to:

$$E_\eta = \frac{1}{\sqrt{H}} \frac{\partial \Phi}{\partial \eta} = \frac{1}{\sqrt{H}} \frac{\Phi_0}{\sinh(\eta) \ln(\coth(0.5\eta_0))} \qquad (15)$$

**Results and Discussion**

We considered the transition process in distilled water when a voltage is applied to the needle-plane electrode system. The following set of parameters were chosen in our calculations: dielectric permittivity $\varepsilon = 81$; $\alpha = 1.5$; a negative pressure threshold at which cavitation starts, -30 MPa; the radius of the electrode tip, $r_{el} = 5$ μm; the eksintrisitet (the ratio of the major semiaxis of the prolate ellipsoid to the small semiaxis) $\xi = 4$.

In Figure 1, the geometry of the problem and the grid in the prolate spheroidal coordinates are presented.

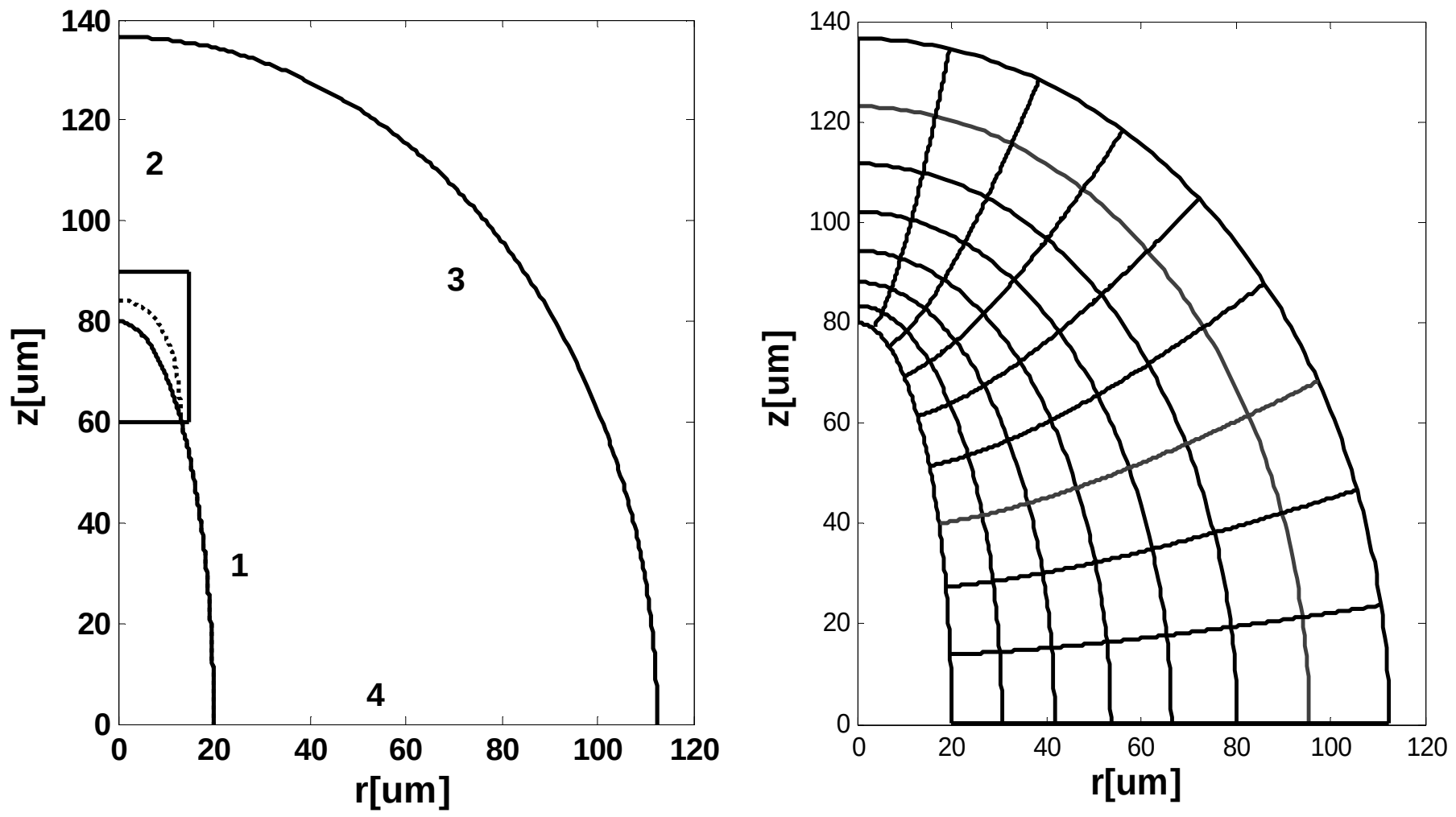

**Fig.1.** The boundaries of the area of integration of equations (5,6) and the grid in prolate spheroidal coordinates: 1 is the electrode surface, 2 is the symmetry axis, 3 and 4 are of the computational domain. For the area marked with rectangle, the two-dimensional distributions of pressure and velocity in Figure 4 are shown. The dashed line schematically shows the region where the absolute value of the electrostrictive pressure exceeds the absolute value of the critical negative pressure necessary for cavitation.

The time dependent sistem of equations (5), (6) together with the equation of state (7) and the boundary conditions (10) was solved in prolate spheroidal coordinates (13) using a McCormack second-order scheme [19].

In all computed cases, the linear form of the voltage pulse $U(t) = U_0 t / t_0, t \le t_0$ was assumed. Here, $U_0 = 7$ kV is the maximal voltage on the electrode, $t_0$ is the front duration. To study the effect of voltage rise time, calculations were performed for $t_0$=1, 5, 10 and 15 ns. A negative pressure in the fluid on the symmetry axis (*r*= 0*, z*), caused by the electrostrictive forces at different moments of the voltage pulse, is shown in Fig. 2.

Electrostrictive forces cause the fluid flow to the electrode. As a result, the absolute value of the total pressure in the fluid is $|p_{tot}| = |p + p_E| < |p_E|$, $p_E = -0.5\alpha\varepsilon\varepsilon_0 E^2$ . Thus, at relatively sharp voltage pulse fronts, conditions arise for the development of cavitation, whereas cavitation under these conditions can not arise at more gentle pulse fronts. This is clearly demonstrated by the computed total pressure distributions on the symmetry axis at different times shown in Fig. 3 (A-D), where the dashed line shows the threshold pressure for cavitation.

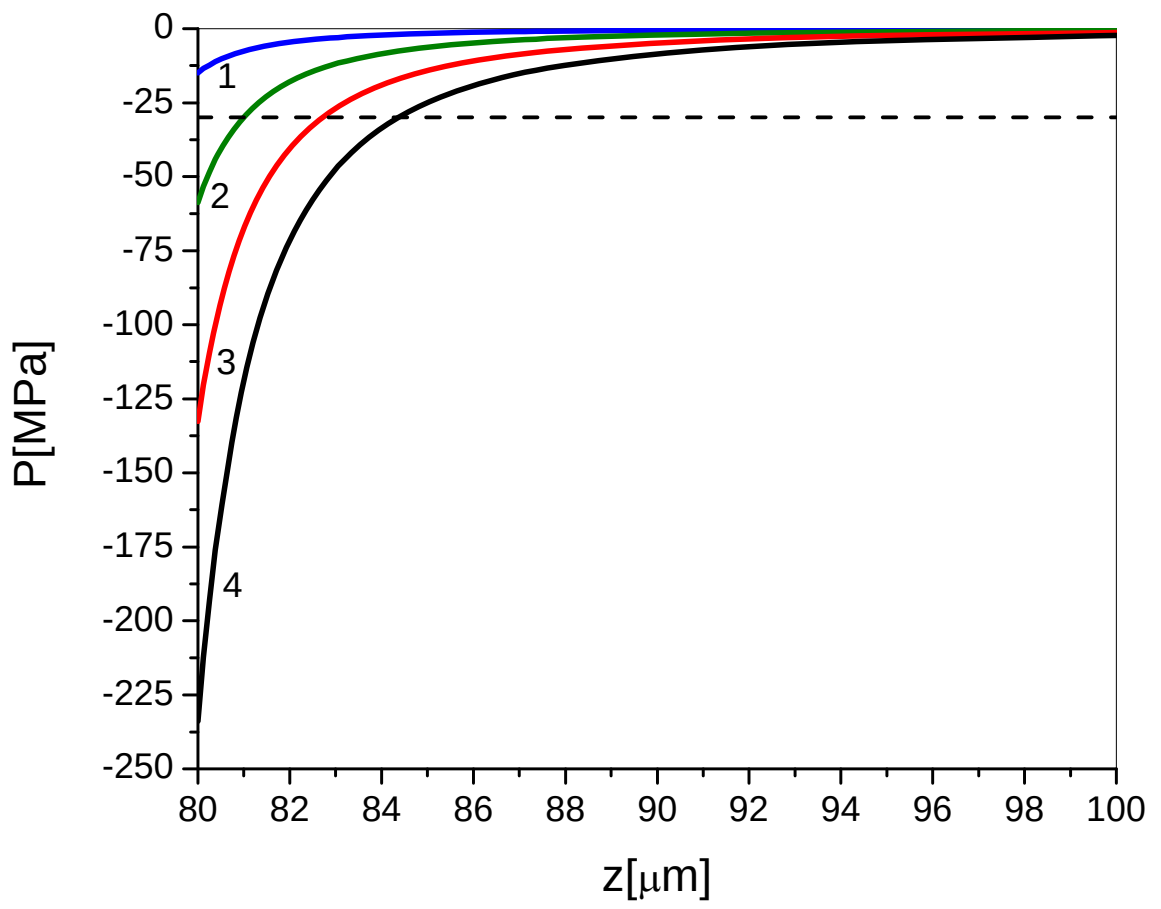

**Fig.2.** (color online) Electrostrictive negative pressure $p_E = -0.5\alpha\varepsilon\varepsilon_0 E^2$ in the fluid along the symmetry axis (r = 0, z) at the time moments: $t/t_0 = 0.25$ (curve 1); 0.5 (curve 2); 0.75 (curve 3) and 1 (curve 4). The dashed line shows the pressure threshold for cavitation when a rupture of continuity of fluid occurs.

In [10], it was shown that the size of the area of the negative pressure, where the conditions for the fluid cavitation raptures are fulfilled, is proportional to the square of the applied voltage amplitude and decreases inversely proportional to the fourth power of the radius of the tip of the needle-like electrode. The performed calculations are in agreement with these qualitative regularities.

Velocity of the fluid flow arising under the considered conditions during the entire voltage pulse remains subsonic and does not exceed tens of m/s (Fig. 3, middle column). Fluid influx to the electrode causes changes in density. However, in all computed cases, the maximum change in fluid density in the vicinity of the electrode does not exceed a few percent (Fig. 3, right column).

The obtained results show a qualitative difference between the behaviors of the liquid at a relatively fast or slow rise of a nonuniform electric field. At a short rise time, there are large tensile stresses (large negative pressure), which can lead to discontinuities and cavitation formation of nanopores. At a relatively slow increasing of the field, the arising flow leads to a strong decrease of the negative pressure down to values below the cavitation threshold, and fluid ruptures do not occur.

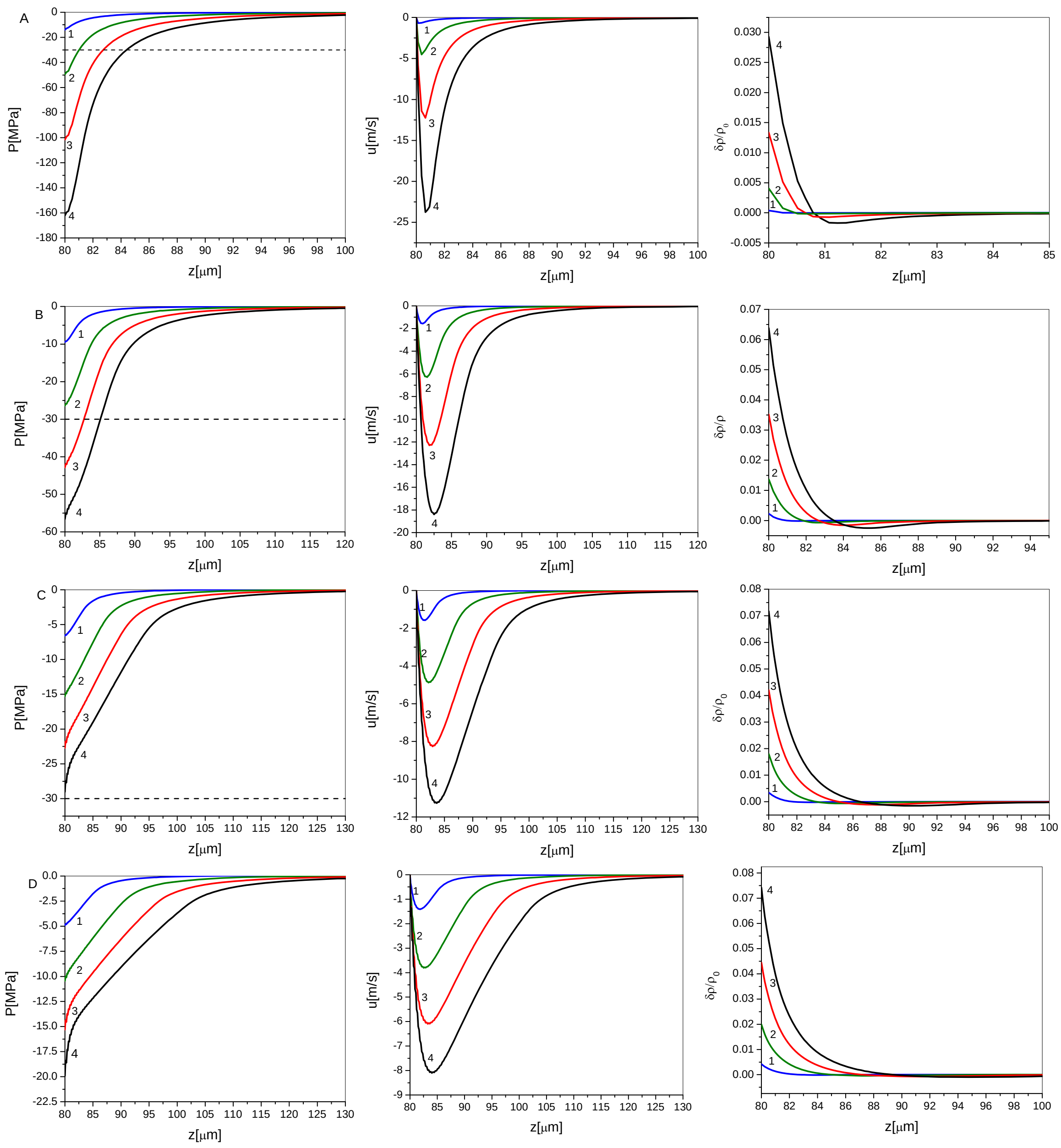

**Fig. 3.** (color online) Longitudinal distributions of the total pressure $p_{tot}$; flow velocity $u_z$ and relative density perturbation $\delta\rho/\rho_0$ along the symmetry axis ($r$=0, $z$). $A$ – for the pulse with the rise time $t_0 = 1$ ns, $B$ – $t_0 = 5$ ns, $C$ - $t_0 = 10$ ns, $D$ - $t_0 = 15$ ns. Curve 1 corresponds to the time moment $t/t_0 = 0.25$, 2 - $t/t_0 = 0.5$, 3 - $t/t_0 = 0.75$, 4 - $t/t_0 = 1$. The dashed line shows the pressure threshold for cavitation when a rupture of continuity of fluid occurs.

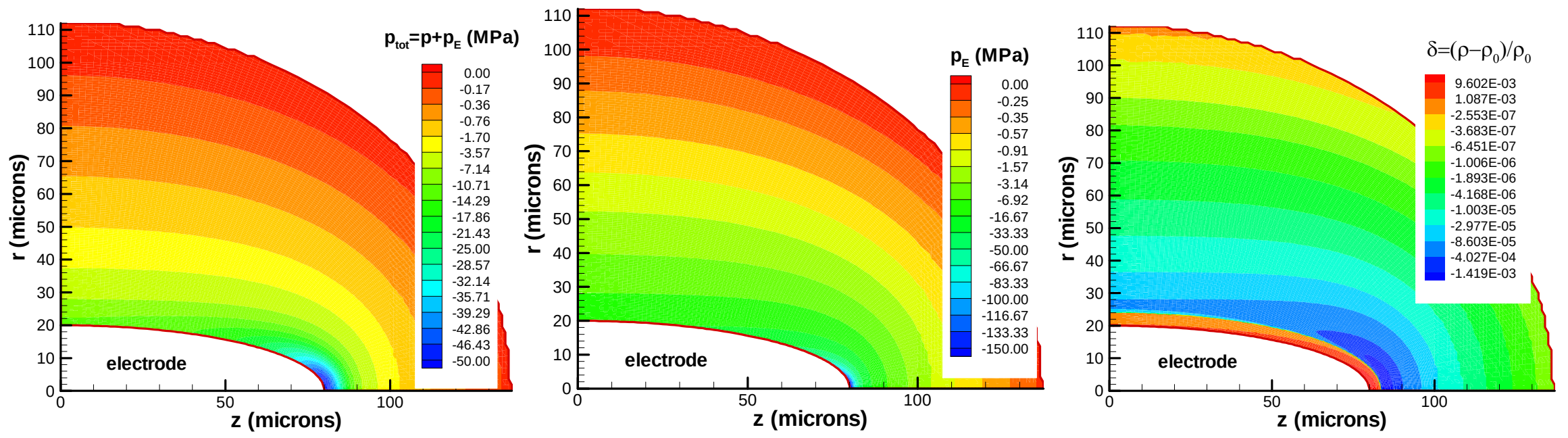

**Fig. 4.** (color online) Contours of the total and ponderomotive pressures, and the relative density perturbation at $t = 5$ ns for a pulse with the front duration $t_0$ = 5ns.

Figure 4 shows the contours of the total, $p_{\text{tot}}$, and ponderomotive, $p_E$, pressures and the relative density, $\delta = (\rho - \rho_0)/\rho_0$, at $t = 5$ ns for a pulse with the front duration $t_0$ = 5ns. The region of the electrode with the absolute value of negative pressures greater than 30 Mpa, where the conditions for the fluid rupture (formation of cavitation micropores) (left figure) is relatively small and extends in the vicinity of the electrode tip to the distance of about 5 μm. The fluid density perturbation (right figure) is maximal in the vicinity of the electrode surface and then, a small region of rarefaction occurs due to fluid motion and stretching under the influence of ponderomotive forces.

Note that for a dielectric fluid with the dielectric constant much lower than that of water, at the same electrode geometry and parameters of the voltage, pulse cavitation ruptures can not be formed. For the reason that the tensile stresses, determined by electrostriction force (8), are linearly dependent on the dielectric constant of the liquid.

## Stretching of Cavitation Micropores Under the Influence of Electrostriction Forces

As already noted, if the negative total pressure is greater than the absolute value of the cavitation threshold, a discontinuity (nanopores) is formed. These nanopores are stretching anisotropically in a nonuniform electric field, mainly in the direction toward the electrode, i.e. in the direction of the electric field gradient., Stretching of nanopores, in the reference frame associated with the fluid, can be described by the theory of elasticity [20]. Here, we restrict ourselves to a simple qualitative analysis.

Consider the nanopore in the fluid in the electric field created by the voltage $U$ applied to the spherical electrode of a small radius $r_{el}$. In this case, the electrostriction force acting on the fluid (2):

$$\vec{F}(R) = -\frac{\varepsilon_0}{2}\frac{\partial\varepsilon}{\partial\rho}\rho\left(Ur_{el}\right)^2\frac{\vec{R}}{R^6}, \qquad (16)$$

where $R$ is the distance from the center electrode to the center of the spherical pores (Fig.5). The difference between the forces acting on the walls of the pore of the radius $b$ in the directions $z$ and $r$ are equal to:

$$F_r \approx \left|\vec{F}(R)\right| \sin\varphi \approx \left|\vec{F}(R)\right| \frac{b}{R} = \alpha\varepsilon_0\varepsilon(Ur_{el})^2 \frac{b}{R^6}$$

$$F_z \approx \left(\left|\vec{F}(R+b/2)\right| - \left|\vec{F}(R-b/2)\right|\right)\cos\varphi \approx \frac{\partial\left|\vec{F}\right|}{\partial R} b = 5\alpha\varepsilon_0\varepsilon(Ur_{el})^2 \frac{b}{R^6} \qquad (17)$$

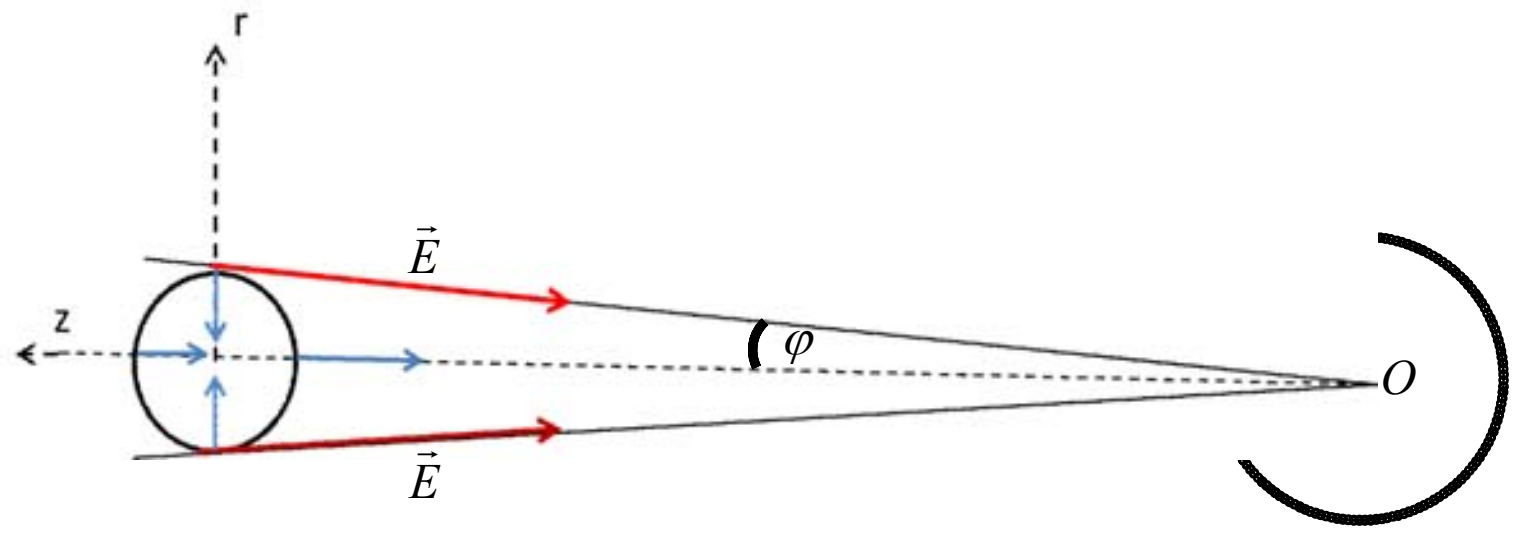

**Fig. 5.** (color online) Directions of the ponderomotive forces acting on the micropore in the vicinity of a small spherical electrode with a center at the point *O*. Vertical arrows show the compressive forces, $F_r$. Force $F_z$, stretching pore in the z direction, is the difference of the forces acting on the front wall of the pore (closer to the electrode) and the rear, located farther from the electrode.

Thus, in the considered case, the tensile strength on the pores along the field is 5 times higher than the force acting in the radial direction. Therefore, under the action of electrostriction forces, the pores will be stretched mostly along the electric field, which is consistent with observations for the bubbles in dielectric liquids [21-25].

Therefore, an intensive formation of pores in the region of negative pressure, which exceeds the cavitation threshold, results in very slight changes of the volume of fluid. Assuming that $l \approx 10$ nm is longitudinal dimension of stretched pores and $b \approx 1$ nm is transversal radius, the density of pores is $n_p \sim \frac{\Delta V}{V}\frac{1}{b^2 l}$. For example, in a case of the density of the pores $n_p \approx 10^3 \mu\text{m}^{-3}$ (in this case the distance between the cavitation pores of the order of 100 nm), the relative changing in volume, $\Delta V / V = 10^{-5}$.

### Flow Arising at Adiabatic Switching of Voltage and its Rapid Shutdown

If the voltage on the electrode is switching slowly enough, the flow occurring in fluid has time to reduce the total pressure to such an extent that the cavitation ruptures can not appear. In this case, the hydrostatic pressure at the electrode can reach the value $p = |p_E|$. At a sharp turning off of the applied voltage, the electrostriction pressure disappears, and a large gradient of the hydrostatic pressure leads to formation of fluid flow from the electrode. As a result, due to inertia of the fluid flow, the formation of negative pressure regions and cavitation ruptures is possible.

Figure 6 shows the formation of negative pressure near the electrode after a sharp (instantaneous) shutdown of the electric field. As the initial condition, the hydrostatic pressure was taken equal to the absolute value of electrostrictive $p = |p_E| = 0.5\alpha\varepsilon\varepsilon_0 E^2$, at the maximum voltage on the electrode $U_0$. It is clearly seen that a region of negative pressure forms in the vicinity of the electrode within a time about a few nanoseconds.

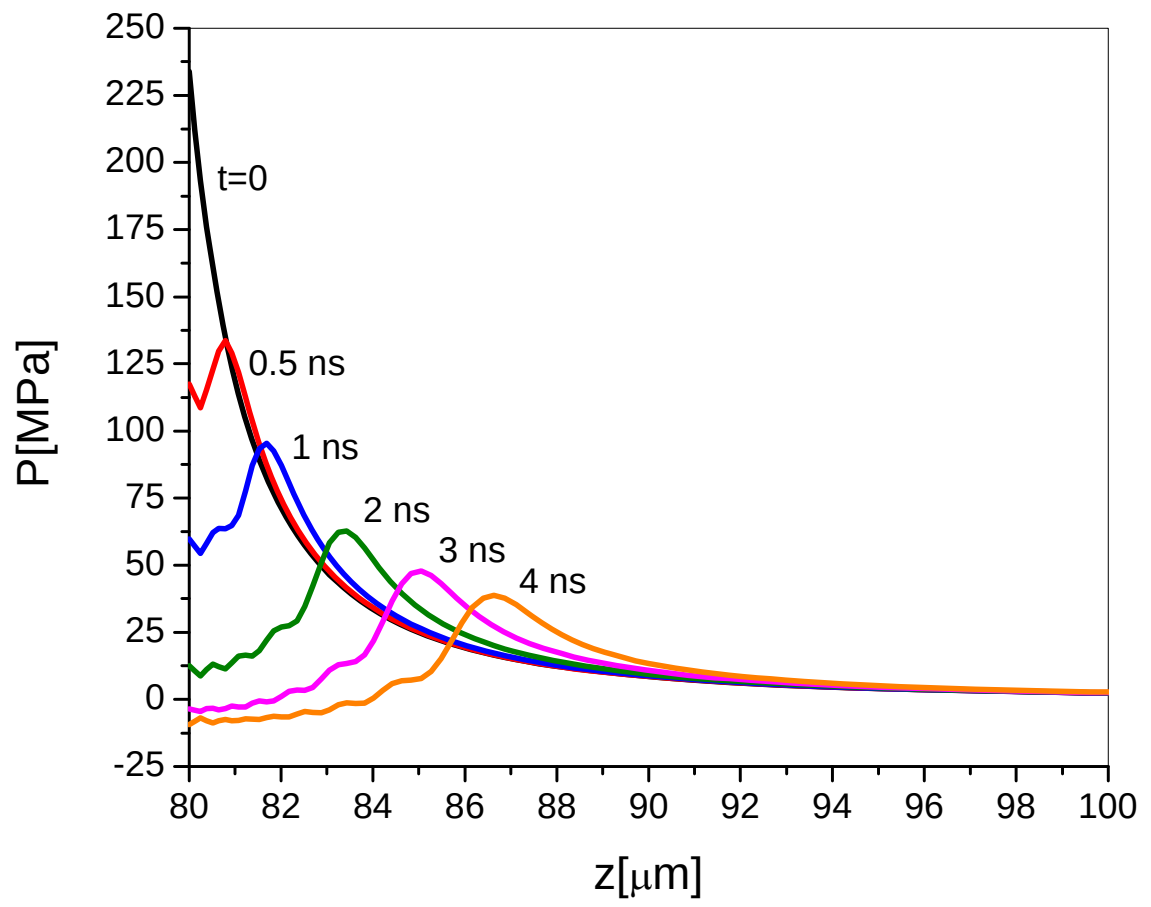

**Fig. 6.** (color online) Longitudinal distributions of the hydrostatic pressure along the symmetry axis ($r$=0, $z$) at different time moments after the voltage interruption.

## Linearized Equations and Example Results

As can be seen from the calculations, changes in the fluid density do not exceed a few percent, and the resulting flow rate is much less than the velocity of sound, so the system of equations (5-7) can be simplified by linearizing it. In the spherically symmetric case, the system of equations (5-7) is reduced to the form:

$$\begin{cases} \dfrac{\partial u}{\partial t} = \dfrac{1}{\rho_0}\dfrac{\partial}{\partial r}\left(p - \dfrac{\alpha}{2}\varepsilon_0\varepsilon E^2\right) \\ \dfrac{\partial p}{\partial t} = \rho_0 c_s^2 \dfrac{1}{r^2}\dfrac{\partial}{\partial r}\left(r^2 v\right) \\ c_s = \sqrt{\dfrac{B\gamma}{\rho_0}} \approx 1500 \text{ m/s} \end{cases} \quad (18)$$

Figure 7 shows the total pressure and fluid velocity for a spherical electrode of radius 5 μm, for voltage pulses with rise 1ns, 5ns, 10ns, 15 ns, obtained by solving the simplified system of equations (18). The voltage amplitude on the electrode was chosen $U_0 \approx 3.32$ kV, that the electric field on a spherical electrode was equal to the field at the end of the ellipsoidal electrode. It shows a good agreement with the results of 2D calculations shown in Fig.4, for the pulses with short fronts (1ns, 5ns) and a reasonable agreement for longer pulses.

This is related to the fact that sizes of the negative pressure region are different in the cases of an ellipsoidal electrode and a spherical electrode.

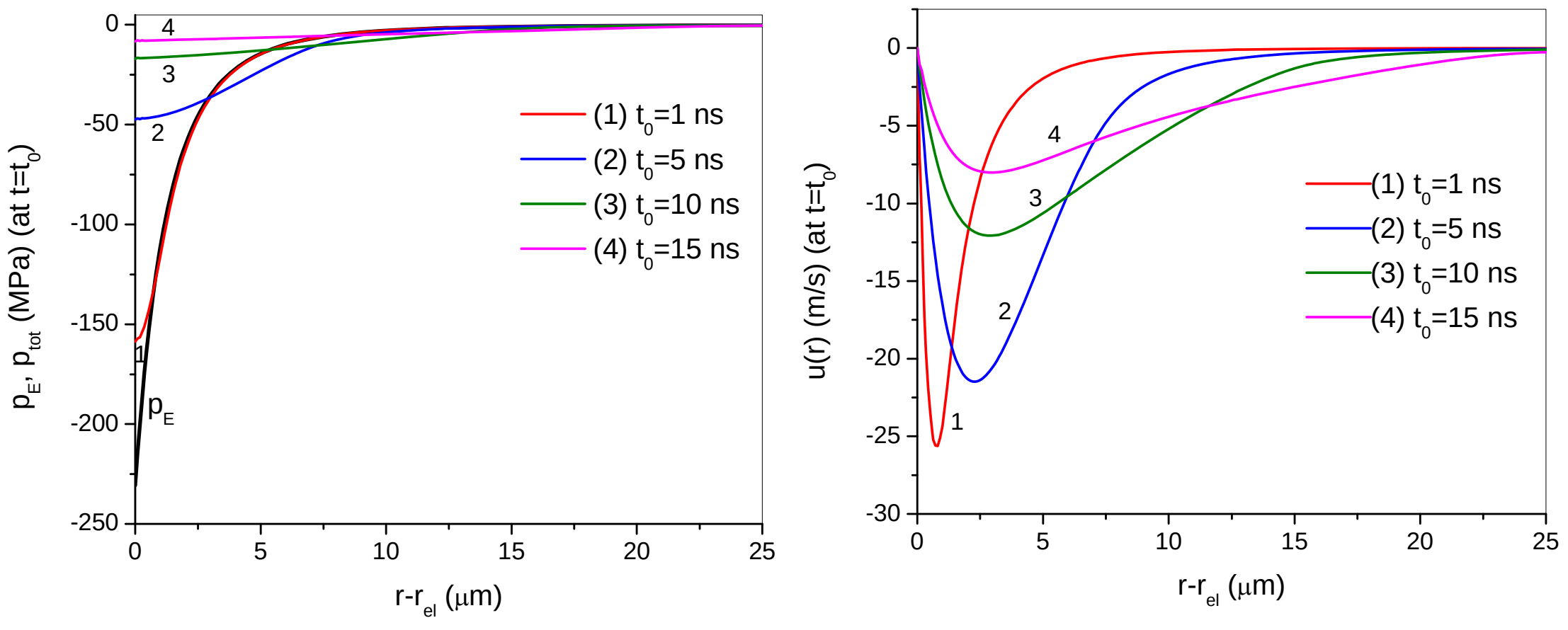

**Fig. 7.** (color online) Radial distributions of electrostriction and the total pressure (left) and the fluid velocity (right) for a spherical electrode of radius 5 μm at the time $t = t_0$ for the voltage pulses with fronts $t_0 =$ 1ns, 5ns, 10ns and 15 ns obtained as a result of calculation of the linearized system (18).

## Conclusions

Pre-breakdown behavior of the dielectric fluid in the pulsed, strong inhomogeneous electric fields was studied. It is shown that in the sub-nano- and nanosecond high-voltage pulses applied to sharp needle-like electrodes, the regions of the negative pressure in the vicinity of the electrode's tip are formed as a result of volumetric electrostrictive forces, which can lead to rupture of fluid due to cavitation.

The cavitation microruptures are strongly extended along the electric field and, even at their significant densities, relative change in volume of fluid is very small.

If the voltage pulse is relatively long, the flow forms in the direction toward the electrode, reducing the total pressure in the fluid. As a result, the negative pressure in the fluid is insufficient for the cavitation, Hence, the micro ruptures of fluid do not occur.

At a relatively slow rising of the applied voltage with subsequent sudden shutdown, a flow of liquid from the electrode is formed with subsequent development of the negative pressure region due to inertia of the fluid.

In all considered cases the direct and reverse flows of fluid induced by the electrostrictive forces are subsonic.

## References

[1]. M. Faraday, Experimental Researches in Electricity (Classic Reprint), (Dover, 2004)

[2] L.D. Landau, E.M. Lifshitz, *Electrodynamics of Continuous Media*, 2nd ed. (Oxford, Pergamon Press, 1984)

[3] J. D. Jackson, *Classical Electrodynamics,* 3rd ed. (Wiley, New York, 1999).

[4] I. E. Tamm*, Fundamentals of The Theory of Electricity* (Moscow: Mir Publ., 1979)

[5] V.P. Skripov, *Metastable liquids* (J. Wiley, New York, 1973).
[6] E. Herbert, S. Balibar and F. Caupin, Phys. Rev. E **74**, 041603 (2006).
[7] V.E. Vinogradov, Technical Phys. Letters., **35**, 54 (2009).
[8] W. An, K. Baumung, H. Bluhm, J. Appl. Phys. **101**, 053302 (2007).
[9] A.Starikovskiy, Y.Yang, Y.I. Cho, A.Fridman, Plasma Sources Sci. Technol. **20,** 024003. (2011).
[10] M.N. Shneider, M. Pekker, A. Fridman, IEEE Transactions on Dielectrics and Electrical Insulation **19**, 1579 (2012).
[11] J.-Z. Zhang, R.K. Chang, Optics Letters, **13**, 916, (1988)
[12] W.K.H. Panofsky and M. Phillips, *Classical Electricity and Magnetism* (Addison-Wesley Pub. Co.,1962).
[13] J.S. Jakobs, A.W. Lawson, J. Chem. Phys. **20**, 1161 (1952).
[14] V.Y. Ushakov, V.F. Klimkin and S.M. Korobeynikov, *Breakdown in liquids at impulse voltage* (NTL, Russia, Tomsk, 2005).
[15] L.D. Landau, E.M. Lifshitz, *Fluid Mechanics,* 2nd ed. (Pergamon Press, 2nd edition, 1987).
[16] Yuan-Hui Li, J. Geophys. Res., **72**, 2665 (1967).
[17] R.I. Nigmatulin, R.Bolotnova, The equation of state of liquid water under static and shock compression, Proceedings VI Zababkhin Scientific Readings, Snezhinsk, Russian Federal Nuclear Center VNIITF, 2003, www.vniitf.ru/rig/konfer/6zst/6zst.htm.
[18] P.M. Morse, H. Feshbach, *Methods of Theoretical Physics, Part I* (New York: McGraw-Hill, p. 661, 1953).
[19] D.A. Anderson, J.C. Tannehill, R.H. Pletcher, *Computational fluid mechanics and heat transfer*. (Hemisphere, New York, 1984)
[20] L.D. Landau, E.M. Lifshitz, *Theory of Elasticity*, 2nd ed. (Oxford, Pergamon Press, 1981).
[21] C.G. Garton, Z.Krasucki, Proc. Roy. Soc. **A280,** 211 (1964)
[22] B.M. Smolyak, Effect of the electric field on surface tension of liquid dielectrics. In book: Thermal properties of liquids and explosive boiling (Sverdlovsk, Ural Sci. Center USSR Academy of Science, 79-84, 1976).
[23] R.S. Allan, S.G. Mason, Proc. Roy. Soc. **A267,** 45 (1962).
[24] S. Torza, R.G. Cox, S.G. Mason, Phil. Trans. Roy. Soc. **A269**, 295 (1971).
[25] B.S. Sommers, J.E. Foster, J. Phys. D: Appl. Phys. **45** 415203 (2012).